\documentclass[conference]{IEEEtran}
\IEEEoverridecommandlockouts
\usepackage{multirow}
\usepackage{cite}
\usepackage{amsmath,amssymb,amsfonts}
\usepackage{algorithm}
\usepackage{url}

\usepackage{breakurl}
\usepackage{mathtools}
\usepackage[noend]{algpseudocode}
\usepackage{graphicx}
\usepackage{tabularx}
\usepackage{balance}
\usepackage{textcomp}
\usepackage{comment}
\usepackage{textcomp}
\usepackage{multirow}
\usepackage{enumitem}
\usepackage{microtype}
\usepackage{booktabs}
\usepackage{graphicx}
\usepackage{textcomp}
\usepackage{threeparttable}
\usepackage{tikz}
\usepackage{xcolor}

\newcommand\crule[3][black]{\textcolor{#1}{\rule{#2}{#3}}}
\definecolor{myyellow}{rgb}{0.867, 0.686, 0.275}
\definecolor{myblue}{rgb}{0.180, 0.337, 0.631}
\definecolor{myred}{rgb}{1, 0, 0.1}
\definecolor{mygreen}{rgb}{0.301, 0.612, 0.440}
\definecolor{mypurple}{rgb}{0.455, 0.443, 0.678}
\definecolor{mygrey}{rgb}{0.302, 0.302, 0.302}
\definecolor{f1green}{rgb}{0.99, 0.537, 0.173}
\definecolor{f1blue}{rgb}{0.36, 0.625, 1.0}
\definecolor{f1yellow}{rgb}{0.867, 0.686, 0.275}
\definecolor{f1orange}{rgb}{0, 0.729, 0.619}

\def\BibTeX{{\rm B\kern-.05em{\sc i\kern-.025em b}\kern-.08em
    T\kern-.1667em\lower.7ex\hbox{E}\kern-.125emX}}
    
\makeatletter
\newcommand\notsotiny{\@setfontsize\notsotiny\@vipt\@viipt}
\makeatother

\begin{document}

\setlength{\textfloatsep}{0pt}

\title{Sparq: A Custom RISC-V Vector Processor for Efficient Sub-Byte Quantized Inference}


\author{\IEEEauthorblockN{Théo Dupuis, Yoan Fournier, MohammadHossein AskariHemmat, Nizar El Zarif,\\  François Leduc-Primeau, Jean Pierre David, Yvon Savaria}
\IEEEauthorblockA{\textit{Department of Electrical Engineering, Polytechnique Montréal, Québec, Canada}}
}
\maketitle


\begin{abstract}
Convolutional Neural Networks (CNNs) are used in a wide range of applications, with full-precision CNNs achieving high accuracy at the expense of portability. Recent progress in quantization techniques has demonstrated that sub-byte Quantized Neural Networks (QNNs) achieve comparable or superior accuracy while significantly reducing the computational cost and memory footprint. However, sub-byte computation on commodity hardware is sub-optimal due to the lack of support for such precision. In this paper, we introduce Sparq, a Sub-byte vector Processor designed for the AcceleRation of QNN inference. This processor is based on a modified version of Ara, an open-source 64-bit RISC-V ``V'' compliant processor. Sparq is implemented in GLOBAL FOUNDRIES 22FDX FD-SOI technology and extends the Instruction Set Architecture (ISA) by adding a new multiply-shift-accumulate instruction to improve sub-byte computation effciency. The floating-point unit is also removed to minimize area and power usage. To demonstrate Sparq performance, we implement an ultra-low-precision (1-bit to 4-bit) vectorized conv2d operation taking advantage of the dedicated hardware. We show that Sparq can significantly accelerate sub-byte computations with respectively 3.2 times, and 1.7 times acceleration over an optimized 16-bit 2D convolution for 2-bit and 4-bit quantization.

\end{abstract}


\section{Introduction}

Convolutional Neural Networks (CNNs) are used in a broad range of applications that include image processing, speech recognition, and natural language processing. As new CNN architectures and optimization methods allow for higher accuracy, the required amount of computations and on/off-chip memory accesses tends to increase. The high complexity and energy consumption pose a challenge for their deployment on resource-constrained devices such as low-power systems. 

In the last few years, research has focused on minimizing computational cost and memory footprint while trying to preserve accuracy. This was achieved through the emergence of multiple new optimization techniques. Among them, reducing the bit-precision of weights and activations allows for a decrease in both the computational complexity and the memory footprint. In extreme cases, where weights and activations are quantized down to 1-bit or 2-bit, the resulting models typically experience an accuracy drop of less than 10\% compared to their full-precision counterparts
\cite{https://doi.org/10.48550/arxiv.1808.00278, https://doi.org/10.48550/arxiv.1603.05279, MLSYS2019_006f52e9,https://doi.org/10.48550/arxiv.1602.02830, BinaryConnect}.

At the same time, new hardware solutions are proposed to address the issue of complexity through the emergence of vectorized Instruction Set Architectures (ISAs). Supporting these new instructions, vector processors can achieve high efficiency through parallelization of tensor computations while maintaining the flexibility of being fully programmable \cite{Cavalcante_2020,Perotti_2022,Stephens_2017}. However, they are limited by the minimum granularity of their vector registers. Generally fixed at 8-bit, this limitation makes these architectures sub-optimal for sub-byte computing.

In this paper, we propose to modify Ara\cite{Cavalcante_2020}, a 64-bit RISC-V RVV1.0 compliant vector processor to add a new vectorized multiply-shift-accumulate instruction, \texttt{vmacsr}. In order to minimize area and power usage, we remove the vector floating-point unit (FPU). The main idea is to increase the performance of ultra-low-precision computations with a cost-effective instruction. To demonstrate the performance of Sparq, we implement several low-precision conv2d algorithms over a wide range of precisions, using the recently proposed ULPPACK \cite{MLSYS2022_14bfa6bb} technique. We demonstrate that the new \texttt{vmacsr} instruction allows us to mitigate the constraints related to the ULPPACK technique and significantly improve performance. Using up to 2-bit quantization, we achieve a speedup of 3.2$\times$ over an optimized 16-bit conv2d implementation and 1.7$\times$ using up to 4-bit quantization. Lastly, we implement the processor in \textsc{Global Foundries 22FDX FD-SOI} technology and compare power and area usage between Sparq and Ara.

\section{Background and Related Work}
\subsection{Sub-byte Convolutional Neural Networks}
Allowing to address both computation complexity and memory footprint, quantization has aroused a real interest. In fact, quantizing properly can achieve considerable memory savings with limited accuracy drop. On Deep Neural Networks such as ResNeXt\cite{8100117} and EfficientNet\cite{tan2020efficientnet}, only a low accuracy degradation was observed when quantizing the full-precision 32-bit weights and activations down to 8-bit\cite{Nagel2021AWP}.
In recent work, ultra-low-precision quantization ($\leq$ 8-bit), which we refer to as sub-byte quantization, has demonstrated that minimal accuracy degradation was achievable with new network structures and training strategies. In some cases sub-byte Quantized Neural Networks (QNNs) perform better than their full-precision counterparts \cite{https://doi.org/10.48550/arxiv.1902.08153, https://doi.org/10.48550/arxiv.2202.09009} as shown in Table~\ref{tab:lsq_quant}.

\begin{table}[htbp]
\vspace{-0.2cm}
\small
\centering
\caption{Accuracy on Quantized Resnet18 Using LG-LSQ~\cite{https://doi.org/10.48550/arxiv.2202.09009}
}
\vspace{-0.3cm}
\begin{tabular}{ccccc}
\toprule
Dataset & Model & Precision (W/A) & Top-1 & Top-5 \\ \midrule
\multirow{4}{*}{ImageNet} & \multirow{4}{*}{Resnet18} & 
LG-LSQ(3/3) & 70.31 & 89.55 \\ \cmidrule{3-5} 
 &  & LG-LSQ(4/4) & 70.78 & 89.77 \\ \cmidrule{3-5} 
 &  & FP32 & 69.76 & 89.08 \\ \bottomrule
\end{tabular}
\label{tab:lsq_quant}
\end{table}

This was made possible by the introduction of architectures and methods focused on ultra-low-precision. LSQ\cite{https://doi.org/10.48550/arxiv.1902.08153}, SSG \cite{https://doi.org/10.48550/arxiv.2202.09009} and DSQ \cite{https://doi.org/10.48550/arxiv.1908.05033} propose to learn the quantization parameters for both weights and activations by minimizing the quantization loss of the network during training. SAWB \cite{MLSYS2019_006f52e9} focuses on the weight quantization parameter using the weight distribution to estimate the optimal quantization scale. To address the issue of unbounded activation range after ReLU, PACT \cite{MLSYS2019_006f52e9} proposes to train a clipping parameter to find the balance point between clipping and quantization error. Focusing on binary or ternary quantization, CNN architectures tailored for 1-bit or 2-bit such as XNORNet\cite{https://doi.org/10.48550/arxiv.1603.05279} or BinaryNet\cite{https://doi.org/10.48550/arxiv.1602.02830} have been introduced. While achieving very high efficiency and low memory footprint, their accuracy suffers a significant drop when compared to full-precision models. All these methods try to minimize the accuracy drop of sub-byte QNN by minimizing the error due to quantization on weights and activations. 
    
\subsection{Sub-byte Computation on Commodity Hardware}

CNNs rely heavily on the use of the \emph{conv2d} operation, which can make up to 90\% of the computation \cite{10.1007/978-3-319-11179-7_36,conv2d_90}. By using ultra-low-precision quantization, or specific binary/ternary architectures, the computational cost can be drastically reduced. Typically, general-purpose processors cannot take advantage of sub-byte operands since their ISA is suited for byte instructions at best. The same applies to their corresponding vector extension ISA whose granularity is typically limited to vectorized 8-bit instructions. Thus, naive implementations of sub-byte algorithms are bounded by the performance of the 8-bit implementation. 

Several acceleration algorithms have been introduced to counterbalance this issue. For example, bit-serial computation allows each operand bit to be processed in a serial manner. As a result, the operation between any arbitrary $N$-bit and $M$-bit precision operand can be computed. However, the computation complexity is defined by $\mathcal{O}(N \times M)$, meaning that this method is only suitable for ultra-low-precision, typically less than 3-bit. For this precision range, the performance over fp32 can be multiplied by a factor of 2 to 6 depending on the precision \cite{auto_kernel}. This paved the way for the development of specialized bit-serial hardware \cite{7783722, https://doi.org/10.48550/arxiv.2302.05996}.

Other techniques, such as ULPPACK, offer a less invasive alternative while achieving higher performance than standard precision implementation. Even though the expected speedup is lower than with bit-serial, typically 2$\times$ to 4$\times$ acceleration over fp32\cite{MLSYS2022_14bfa6bb}, this method covers a wider precision range, typically 1-bit to 4-bit, by utilizing densely packed operands with commodity Single Instruction, Multiple Data (SIMD) architectures. Although no specialized hardware is required, this algorithm can benefit from specialized instruction such as a multiply-shift-accumulate.

\begin{figure}[!h] 
  \centering
    \noindent\includegraphics[width=9cm]{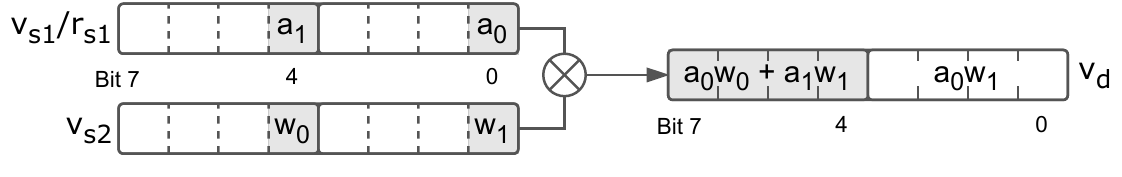}
\vspace{-0.6cm}
\caption{Multiplication between 2 packed 8-bit vectors with 1-bit precision on weights and activation. The dot product of the two packed elements is computed using a single multiplication on the MSB. The result then needs to be shifted to retrieve the dot product.}
\label{fig:dot-prod}
\end{figure}

\section{Software implementations}
In this section, we provide an overview of different precision \emph{conv2d} algorithms benchmarked on Sparq. Every implementation, except for the provided benchmarks, is handwritten using inline assembly, unrolled, and stored using a channel-first memory layout for the input, kernel, and output tensors.
\subsection{Optimized Vector Conv2d }
\label{CONV2D}

A wide range of benchmarks is provided with Ara\footnote{See https://github.com/pulp-platform/ara}, including a double-precision (DP) \emph{conv2d} function. The DP implementation achieves high utilization of Ara's lanes by using the available optimized vector slides to maximize efficiency through high data reuse. We implemented int16 and fp32 \emph{conv2d} based on the structure of the convolution example to serve as the baseline for our comparisons. The choice of a dedicated convolution algorithm over an image to column (\emph{im2col}) operation followed by a GEneral Matrix Multiplication (GEMM) technique is motivated by the reduction of the memory footprint induced by the \emph{im2col} operation. Thus, fewer on/off-chip memory accesses are required to compute the output. We benchmarked the int16 \emph{conv2d} on Sparq and the fp32 version on Ara. Our implementations of int16 and fp32 achieve a lane utilization of respectively $93.8\%$ and $ 93.6\%$ at $1\times32\times 512\times512$ input size.

\subsection{ULPPACK}
\label{ULPPACK}

ULPPACK \cite{MLSYS2022_14bfa6bb} is a software-only technique aiming towards acceleration of ultra-low-precision computation via an effective packing operand scheme. As illustrated in Figure~\ref{fig:dot-prod} with an 8-bit register example, by packing multiple low-precision operands in wider registers, a single multiplication instruction completes the dot product between multiple operands and returns the result on the 4 most significant bits. Formally, the result can be expressed as follows:
\begin{gather*}
    (a_0 + 2^4 a_1)\times (w_1 + 2^4 w_0) \\ = 2^8 a_1 w_0 + 2^4 (a_0 w_0 + a_1 w_1) + a_0 w_1 
\end{gather*}
    
The dot product of packed vectors has to be extracted through a right logical shift (LSR) before accumulation. Local accumulations of the non-shifted result can be done to alleviate the number of shift operations, but are limited by the width of the low-precision result. Moreover, increasing the precision of weights and activations results in a reduction in the number of possible local accumulations due to overflow, implying a lower speedup. In the Figure~\ref{fig:dot-prod} example, using 8-bit elements to compute the product of 2 packed operands limits the result on a 4-bit width. By using 1-bit precision on weights and activation, 8 local accumulations are possible without risking overflow. This method is an efficient way to compute vector-based algorithms since it computes the dot product between multiple low-precision operands with a unique higher precision multiplication, which significantly reduces the computational cost and resource usage.

\begin{figure}[!h] 
  \centering
    \noindent\includegraphics[width=\linewidth]{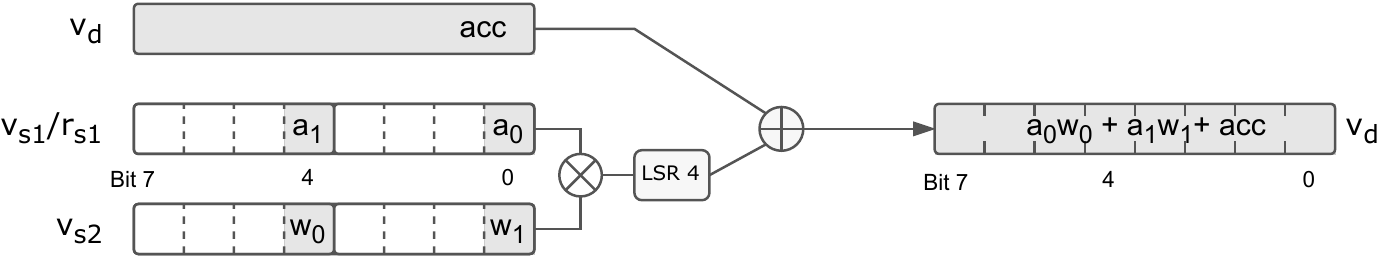}
\vspace{-0.6cm}
\caption{Diagram of the multiply-shift-accumulate operation on 8-bit packed registers with 1-bit precision for activations and weights. A shifter is inserted between the multiplication and the accumulation.}
\label{fig:maccsr}
\end{figure}

\section{Proposed Architecture}

\subsection{\texttt{vmacsr} Custom Instruction}
To address the issue of local accumulation overflow described in Section~\ref{ULPPACK}, we propose to implement a multiply-shift-accumulate operation as a custom vector instruction. As depicted in Figure~\ref{fig:maccsr}, using the multiply-shift-accumulate operation, the low-precision multiplication is shifted and accumulated. This avoids the need for an intermediate register, a logical right shift, and an addition. 

\begin{figure}[!h] 
  \centering
    \noindent\includegraphics[width=\linewidth]{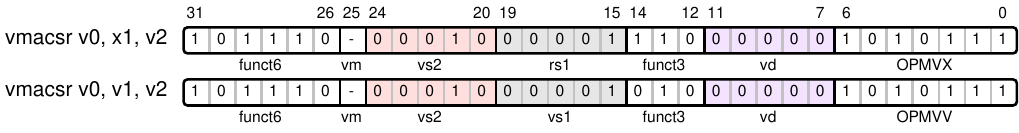}
\vspace{-0.8cm}
\caption{\texttt{vmacsr} encoding example.}
\label{fig:encoding}
\end{figure}

In this work, we mainly focus on a specific precision range following the condition $N + M \leq 7$-bit, where $N$ and $M$ are the operand's precision in bits. In this range, quantized models offer sufficient precision to achieve comparable or superior accuracy compared to full-precision models. We implement a \emph{conv2d} with 16-bit packed register granularity to achieve high performance on this range. However, the use of multiply-shift-accumulate is applicable on higher precision operands and elements width, addressing the same overflow issue. To implement the new instruction, we modified Ara's dispatcher to specify the \texttt{vmacsr funct6} encoding. As our instruction behaves very closely to \texttt{vmacc}, we use the free \texttt{funct6} encoding following \texttt{vmacc}'s \texttt{funct6} \cite{riscv-v-spec}. We implement \texttt{vmacsr} in both \texttt{OPMVV} and \texttt{OPMVX} format allowing vector-vector and vector-scalar instructions to be used. The encoding of \texttt{vmacsr} is described in Figure~\ref{fig:encoding}. It should be noted that for this work, only the vector-scalar \texttt{vmacsr} instructions is used. We then modified Ara's SIMD multiplier to describe the expected behavior of the instruction, denoted as follows:

\begin{center}
\texttt{$V_d \gets V_d + ((V_{s1} \times V_{s2}) >> M)$}
\end{center}

\noindent
where $M$ is the shifted value on the resulting product before accumulation. In this work, we only pack two operands per register to reduce the overhead related to the packing operation carried out at runtime. Thus, $M$ is hard-wired at half the granularity of the vector registers.

\begin{algorithm}[ht]
\small
\begin{tabularx}{\textwidth}{l>{$}l<{$}X}
$H, C$ &\text{: Input height, channels}\\
$F_h, F_w$ &\text{: Kernel height and width}\\
$M$ &\text{: Packed operand per element}\\
$R, V$ &\text{: Scalar and vector registers (initialized to zero)} \\
\end{tabularx}
\caption{Proposed vector conv2d using \texttt{vmacsr}}\label{alg:cap}
\begin{algorithmic}[1]
\For{$ h \gets 1$ to $H$}
    \State $V_{F_{h}} \gets 0$
    \For{$c \gets 1$ to $\frac{C}{M}$}
        \State $V_0\gets $ load one packed input row \label{input_ld}
        \For{$i \gets 1$ to $F_w$}
            \State $R_{[1:F_h]} \gets $ load the $i^{th}$ packed kernel column \label{kernel_col_ld}
            \For{$j \gets 1$ to $F_h$}
                \State $V_{j} \gets
                \texttt{vmacsr}(R_j, V_0, V_{j})$ \label{vmacsr_op}
            \EndFor
            \State $V_0 \gets \texttt{vslidedown}(V_0, 1)$ \label{slidedown}
        \EndFor
    \EndFor    
    \If{$h \geq F_h $} \label{first_out_row}
        \State $O[h]\gets V_1$ \Comment{store one output row} \label{store}
        \For{$j \gets 1$ to $F_h - 1$}
            \State $V_j \gets V_{j+1}$ \label{move}
        \EndFor
    \EndIf
\EndFor
\end{algorithmic}
\end{algorithm}

\subsection{2D Convolution Algorithm}

To our knowledge, the ULPPACK method has only been implemented on an ARM CPU using the ARM Neon intrinsic ISA extension \cite{Stephens_2017,MLSYS2022_14bfa6bb}. In addition, by rearranging the input and filters with an \emph{im2col} operation, the convolution is performed with a GEMM. However, Ara's dedicated \emph{conv2d} algorithm takes advantage of the \texttt{vslidedown} instruction provided by the RISC-V ``V'' ISA extension, which does not have any ARM equivalent. Thereby, the output stationary ULPPACK \emph{conv2d} algorithm presented in Algorithm \ref{alg:cap} is based on the \emph{conv2d} described in Section~\ref{CONV2D}. Although the operand packing operation is not detailed in the algorithm, it is carried out at runtime. We use the ULPPACK P1 packing \cite{MLSYS2022_14bfa6bb} scheme to compute the contribution of $M$ channels at each iteration, where $M$ is the number of operands packed per register. Following the example in Figure~\ref{fig:dot-prod},  $a_i$ and $w_j$ respectively represent activation and weight values from the channel $i$ and $j$. \\

Algorithm \ref{alg:cap} presents a simplified ultra-low-precision \emph{conv2d} implementation using the \texttt{vmacsr} instruction available on Sparq. For each loaded packed input row (line \ref{input_ld}), the algorithm performs a vector multiply-shift-accumulate operation (line \ref{vmacsr_op}) to compute the partial result with the first packed kernel column. The input is then slided by one element to the left (line \ref{slidedown}) to accommodate the next kernel column, and this operation is repeated until all the kernel columns and input channels have been processed. In order to complete and store the first output row (line \ref{store}), $F_h$ packed input rows must be processed (line \ref{first_out_row}). The partial results contained in the following vector registers (line \ref{move}) are moved down to be completed one by one with each new packed input row.

\begin{figure}[!h] 
  \centering
    \noindent\includegraphics[width=\linewidth]{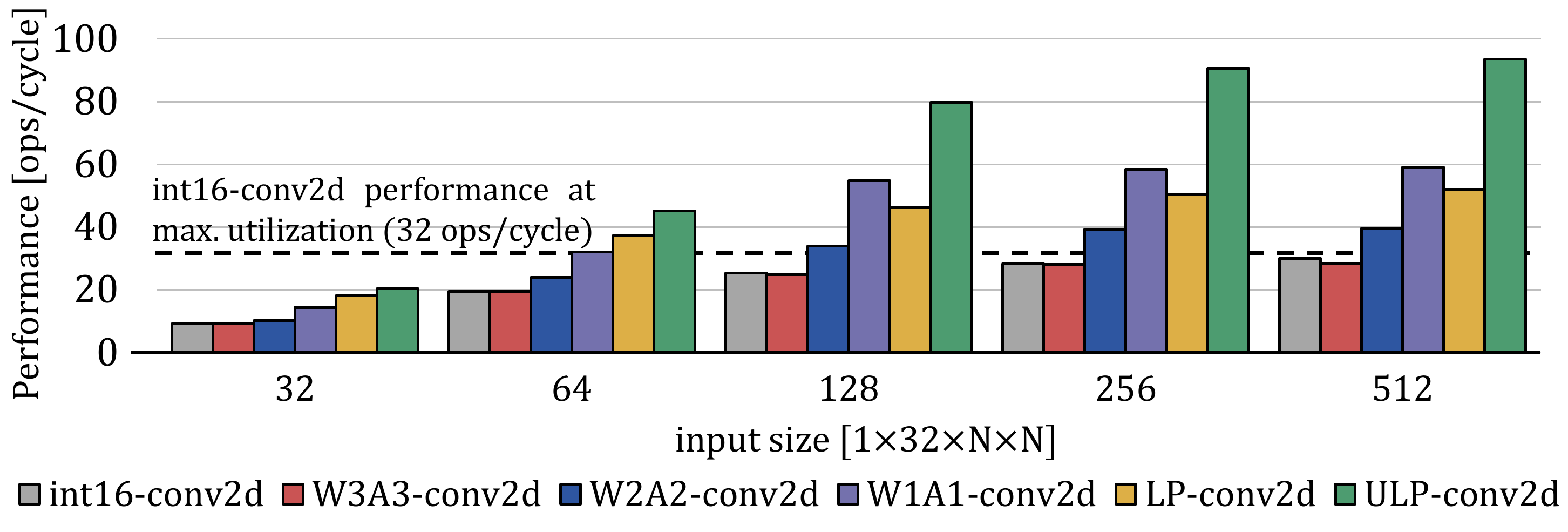}
\vspace{-0.7cm}
\caption{Performance comparison between several \emph{conv2d} implementation using a $7\times7$ kernel size. Ultra-Low-Precision (ULP) and Low-Precision (LP) \emph{conv2d} take advantage of the \texttt{vmacsr} instruction on Sparq while W1A1, W2A2, W3A3 \emph{conv2d} run on native RISC-V ``V" ISA.\\}
\label{fig:results_perf}
\end{figure}

\section{Results and Performance Analysis}
\subsection{Performance Analysis}
\begin{figure}[ht]
\centering
\begin{minipage}{0.48\linewidth}
\includegraphics[height=4cm]{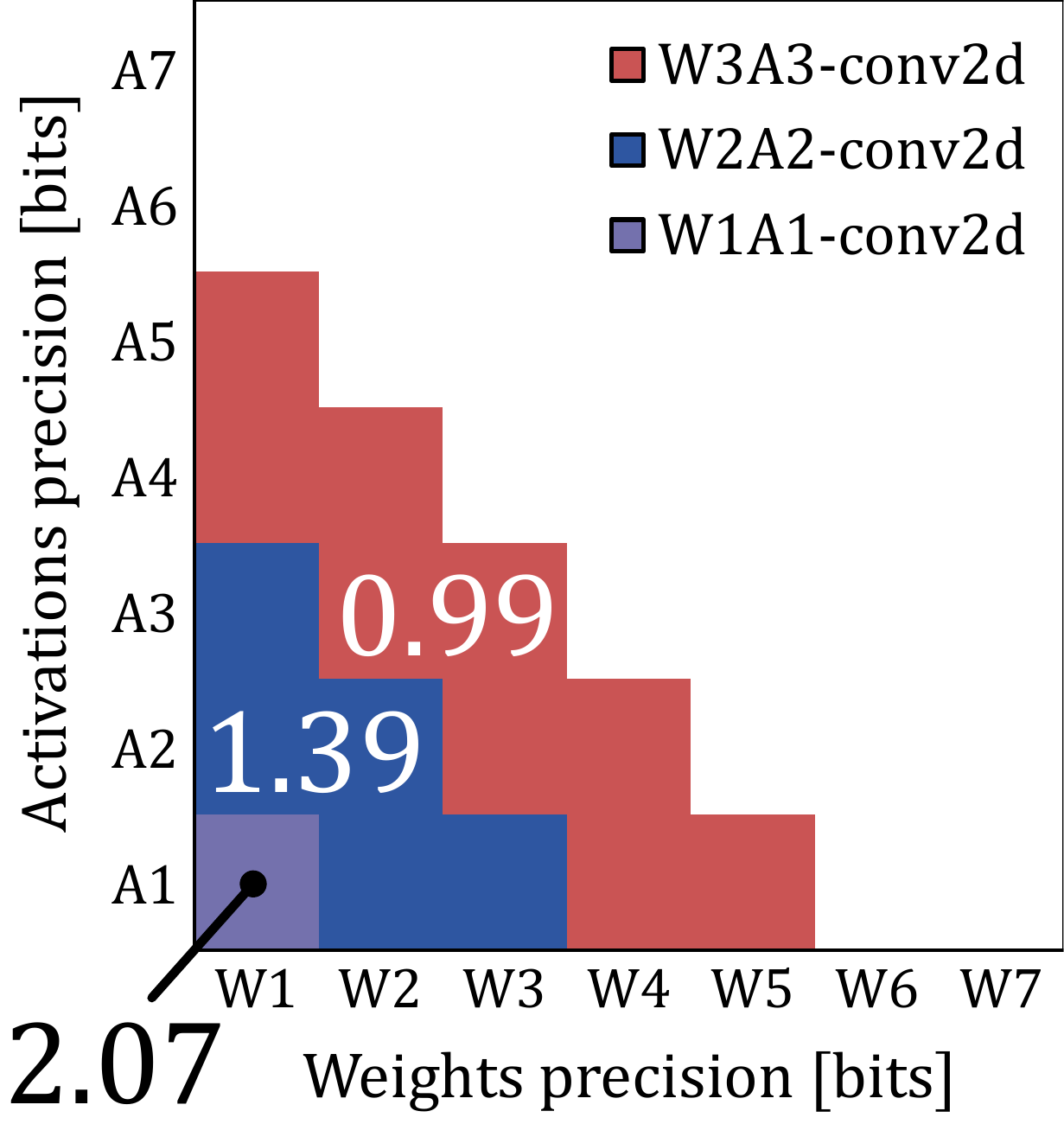}
\label{fig:ara_perf}
\vspace{0.3cm}
\end{minipage}
\hfill
\begin{minipage}{0.50\linewidth}
\centering
\includegraphics[height=4cm]{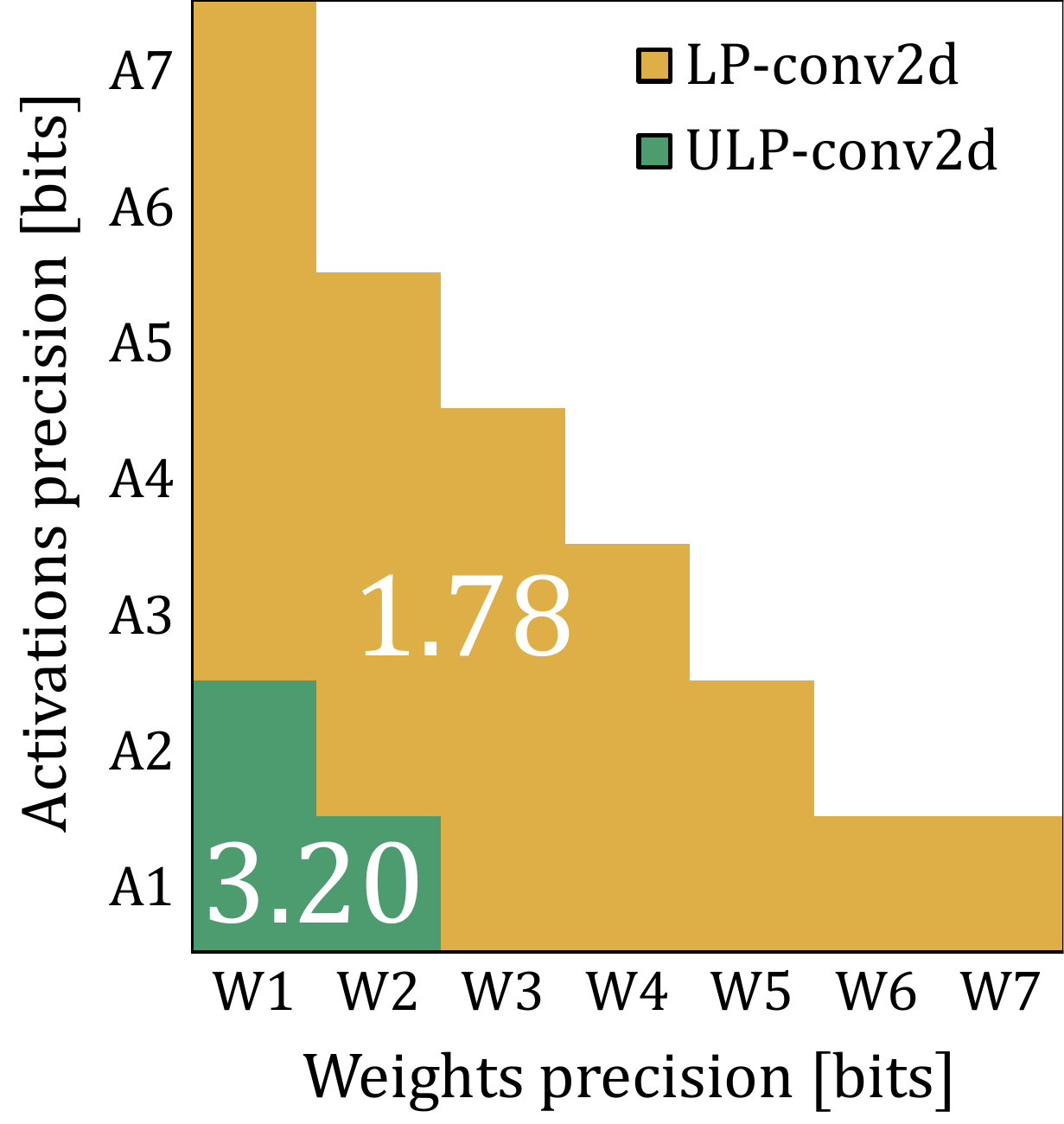}
\label{fig:sparq_perf}
\vspace{0.3cm}
\end{minipage}

\begin{tabular}{cc}
\hspace{0.1cm} \small (a) Native implementation & \hspace{0.5cm} (b) \small \texttt{vmacsr} implementation
    \end{tabular}
\vspace{-0.6cm}
\caption{Relative speedup over the 16-bit \emph{conv2d} implementation on the overflow-free precision region. The kernel size is $7\times7$ over a $32\times256\times256$ input. (a) for the native implementation benchmarked on Ara (b) for the accelerated version taking advantage of the \texttt{vmacsr} instruction benchmarked on Sparq.}
\label{fig:results_precision}
\end{figure}

We compare the performance of the different \emph{conv2d} implementations over several precisions for activations and weights using RTL simulations with a 4-lane configuration for both Ara and Sparq. Our measured execution time includes both activations and weights packing done at runtime. However, the overhead induced by weights packing could be avoided by offline preprocessing. Figure~\ref{fig:results_perf} presents the performance in operations per cycle over the different \emph{conv2d} implementations. \\
As expected, the performance of ULPPACK running on Ara shows an improvement over its 16-bit counterpart. Using $7\times7$ favors high data reuse, hence the small gap with the theoretical throughput. Moreover, using Sparq's \texttt{vmacsr} instruction removes the constraint of local accumulation, offering two major benefits: 

\begin{itemize}[leftmargin=*]
    \item \textbf{Performance increase} resulting from the expected overall reduction in the number of instructions to compute the output matrix as presented in Figure~\ref{fig:results_perf}.
    \item \textbf{Higher precision range}, as shown in Figure~\ref{fig:results_precision}, is obtained without modifying the algorithm. The range is only limited by the 8-bit wide dot product result when using 16-bit packed registers for Low-Precision (LP) and the 4-bit result for the Ultra-Low-Precision (ULP).
\end{itemize}
\vspace{-0.3cm}

\vspace{-0.2cm}
\subsection{Physical Implementation} 
\begin{figure}[ht]
\centering
\begin{minipage}{0.1\textwidth}
\centering
\includegraphics[height=4.5cm]{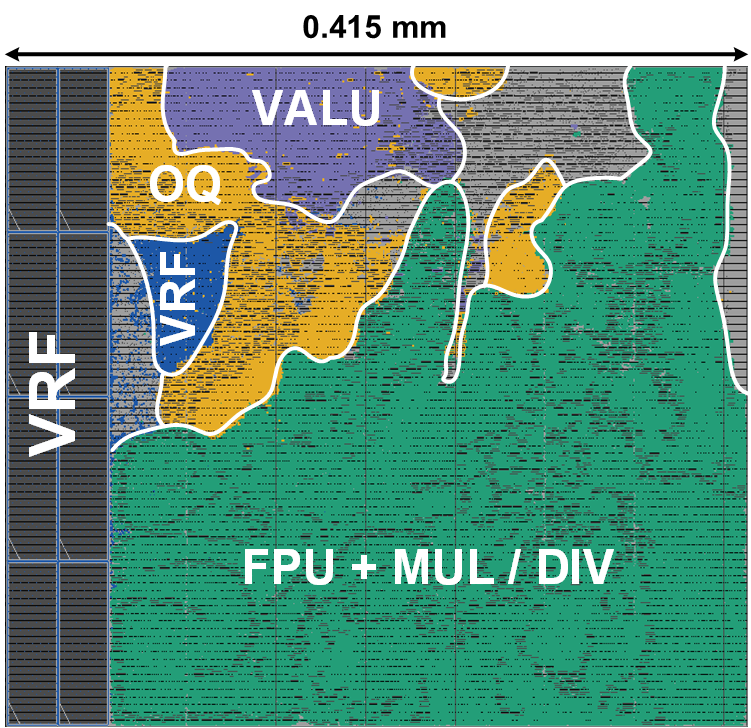}
\label{fig:ara}
\end{minipage}
\hfill
\begin{minipage}{0.19\textwidth}
\vspace{0.1cm}
\centering
\includegraphics[height=4.46cm]{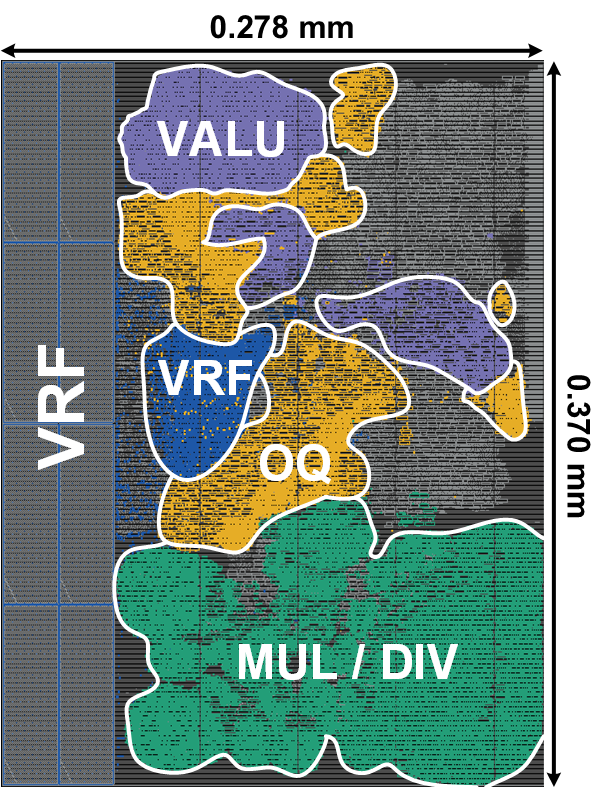}
\label{fig:sparq}
\end{minipage}
\begin{tabular}{cc}\hspace{0.7cm} 
\small (a) Ara lane & \hspace{2.2cm} \small (b) Sparq lane
    \end{tabular}
\vspace{-0.2cm}
\caption{Ara and Sparq lanes placed and routed designs.
         \crule[myblue]{0.2cm}{0.2cm} is the vector register file, 
         \crule[myyellow]{0.2cm}{0.2cm} is the operands queue, 
         \crule[mygreen]{0.2cm}{0.2cm} is the vector fixed point multiplication and division units with or without the FPU,
         \crule[mypurple]{0.2cm}{0.2cm} is the vector ALU.}
\label{fig:pnr_ara_sparq}
\end{figure}

Since the \texttt{vmacsr} circuit is located in the vector engine, we only implemented one lane of Ara and Sparq using \textsc{GlobalFoundries 22FDX FD-SOI} technology. For synthesis and back-end, we used the \textsc{Synopsys Design Compiler S-2021.06-SP5} and \textsc{Cadence Innovus v21.15}, respectively. The physical implementation results of these lanes are summarized in Table~\ref{tab:impl}. As anticipated, the Sparq lane exhibits significant improvements in terms of area (-43.3\%) and power consumption (-58.8\%) when compared to a standard Ara lane, primarily due to the FPU removal. Furthermore, the inclusion of the \texttt{vmacsr} instruction did not impact the typical corner frequency, indicating that it did not affect the critical path of the design. In fact, the removal of the FPU actually yielded an increase (+8.7\%) in the maximum lane clock speed. It is important to note that in Ara, the critical path is primarily located in the VLSU and SLDU, both of which are part of the interconnection between lanes \cite{Cavalcante_2020}. These interconnections were not implemented as part of this work since the addition of \texttt{vmacsr} would have had a negligible impact on those modules. Figure~\ref{fig:pnr_ara_sparq} shows the physical layout of both designs.
\vspace{-0.3cm}
\vspace{-0.5cm}

\begin{table}[ht]
\setlength{\intextsep}{0pt}
    \centering
    \caption{Physical Implementation of Ara and Sparq Lanes}
    \vspace{-0.3cm}
    \begin{tabular}{rl@{\hskip 2em}ll}
    \toprule
     & Ara Lane & \multicolumn{2}{l}{Sparq Lane} \\\cmidrule(lr){2-2} \cmidrule{3-4}
    Number of Lanes & 4 &  4 \\ 
    VRF Size [KiB] & 16 &  16 \\ 
    Lane Cell Area [mm$^2$] & 0.120  & 0.068  \\ 
    Lane Core Frequency\tnote{*} [GHz]     & 1.346  & 1.464       \\
    Lane Power\tnote{*} [mW] & 159.2   & 65.6 \\ \bottomrule
    \end{tabular}
    
    \begin{tablenotes}[normal,flushleft]
        \footnotesize
        \item [*] At typical corner (TT/0.8V/25$^{\circ}$C)
    \end{tablenotes}
\label{tab:impl}
\end{table}
\vspace{-0.3cm}

\section{Conclusion}
This paper introduced the extension of the RISC-V ``V'' ISA with a multiply-shift-accumulate custom instruction, allowing better performance of sub-byte computations. We presented Sparq, a FPU-free RISC-V vector processor supporting the multiply-shift-accumulate instruction. To demonstrate its improved performance, we implemented a \emph{conv2d} algorithm for 1 to 4-bit precision operands. In this range, quantized models offer sufficient precision to achieve comparable or superior levels of accuracy compared to full-precision models. We showed that, with \texttt{vmacsr}, we achieved up to 1.7$\times$ and 3.2$\times$ speedup over an optimized 16-bit \emph{conv2d} depending on the precision. Lastly, the implementation reports of the modified version in GF22 nm showed that removing the FPU significantly reduced the area and power usage, while the newly added instruction had no impact on the critical path. 

In future work, we plan to improve the flexibility of \texttt{vmacsr} using a runtime configurable shifter, as well as testing the proposed algorithm on an FPGA emulation of Sparq.

\section{Acknowledgments}
The authors thank CMC Microsystems and Global Foundries for access to design tools and technologies. This research was funded by CMC Microsystems, Mitacs and the OpenHW
Group.

\clearpage
\bibliographystyle{IEEEtran}
\IEEEtriggeratref{10}
\bibliography{IEEEabrv,refs}

\end{document}